# Orientation-Dependent Handedness and Chiral Design

Efi Efrati[*] and William T. M. Irvine

*James Franck Institute, The University of Chicago, 929 East 57 Street, Chicago, Illinois 60637, USA*
(Received 13 August 2013; revised manuscript received 25 October 2013; published 16 January 2014)

Chirality occupies a central role in fields ranging from biological self-assembly to the design of optical metamaterials. The definition of chirality, as given by Lord Kelvin, associates chirality with the lack of mirror symmetry: the inability to superpose an object on its mirror image. While this definition has guided the classification of chiral objects for over a century, the *quantification* of handed phenomena based on this definition has proven elusive, if not impossible, as manifest in the paradox of chiral connectedness. In this work, we put forward a quantification scheme in which the handedness of an object depends on the direction in which it is viewed. While consistent with familiar chiral notions, such as the right-hand rule, this framework allows objects to be simultaneously right and left handed. We demonstrate this orientation dependence in three different systems—a biomimetic elastic bilayer, a chiral propeller, and optical metamaterial—and find quantitative agreement with chirality pseudotensors whose form we explicitly compute. The use of this approach resolves the existing paradoxes and naturally enables the design of handed metamaterials from symmetry principles.

## I. INTRODUCTION

Determining if a given screw or a twisted fusilli is right or left handed can be done with very little difficulty. However, the general quantification of handedness has proven to be an elusive task [1]. Such a quantification is of exceptional importance in the self-assembly of handed objects [2,3] and in predicting the handed response of natural and manmade materials [4–6]. In 1893, Lord Kelvin coined the term "chirality" to describe the inability to superpose an object on its mirror image [7], referred to here as a lack of mirror symmetry [8]. Associating a number with the handedness of an object such as to be consistent with Kelvin's definition requires that this number changes sign under reflections and reads zero only for objects possessing mirror symmetry. It is now well understood that in a general setting no such quantification is possible [1,9,10]. This is because it is always possible to continuously deform a body into its mirror image without passing through a configuration that is mirror symmetric. This property, known as chiral connectedness, holds for all three-dimensional bodies possessing 5 or more degrees of freedom, in particular, all continuous bodies.

While the term "chirality" comes from the Greek word for "hand" and is often assumed to be synonymous with handedness, some authors draw a distinction between chirality and handedness [11,12] or use different terms altogether to distinguish different notions of chiral behavior, e.g., intrinsic chirality and extrinsic chirality [4,13]. In the present manuscript, we refer to the term chirality, in accordance with Kelvin's definition, as the lack of mirror symmetry and use the term handedness (as distinct from chirality) to describe the property that distinguishes between right- and left-handed objects.

In this work, we interpret the notion of handedness as a relation between directions and rotations, consistent with various handedness manifestations such as thread handedness for a screw, the right-hand rule, R/S [14] classification of organic molecules, and optical activity. We show that this interpretation leads to naturally tensorial and thus orientation-dependent measures of handedness. While isotropic averages of these measures recover Kelvin's definition of chirality, the full measures predict orientation-dependent handed behavior even for objects that are mirror symmetric. This orientation-dependent approach is made quantitative, applied to experiments, and shown to provide a natural tool for the design of metamaterials.

## II. ORIENTATION-DEPENDENT HANDEDNESS

Mirror-symmetric structures can, in some cases, support handed phenomena [4,5,15–18]. For example, the mirror-symmetric crystal of silver gallium sulfide was shown to exhibit optical activity when illuminated along specific directions [19]. In what follows, we demonstrate that an orientation-dependent handedness property can be attributed to such mirror-symmetric systems and that this notion of handedness allows an intuitive and robust interpretation. This serves as a starting point for the attribution of orientation-dependent handedness to general (non-mirror-symmetric) systems.

*efrati@uchicago.edu

Figure 1 shows a thin elastic bilayer whose internal structure is homogeneous in the plane and symmetric under reflections. When long and narrow strips are cut from the bilayer, they curve to form helicoidal strips of both right and left handedness, depending on the relative orientation of the strips and the directions in which the layers were stretched. The handedness of each of the helicoidal strips is easily determined by following the surface's face with the right hand. If advancing along the helicoid's length requires the hand to roll outward, the helicoid is said to be right handed (see, for example, the helicoidal strip in part IV of Fig. 1(c)).

Strips whose long direction is oriented along the $x$ axis form right-handed helicoids, whereas strips whose long direction is oriented along the $y$ axis form left-handed helicoids. The symmetric square cutout [part II of

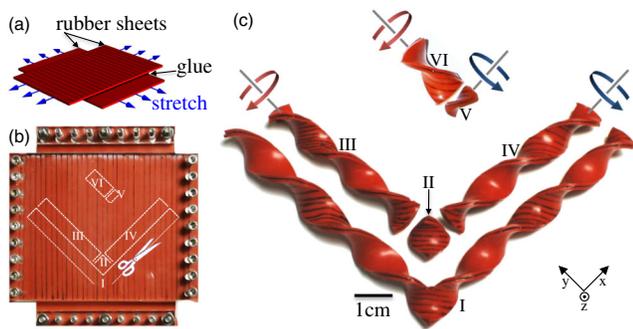

FIG. 1. Orientation-dependent manifestation of handedness in a reflection symmetric continuum. Two identical rubber sheets are uniaxially stretched and glued together to form a rubber bilayer, as done in Ref. [27]. Narrow strips cut from the bilayer curve out of plane to accommodate the difference in rest length between the layers and form helical structures. The square boundaries in II of panel (b) give rise to a cutout [II of panel (c)], which is symmetric under reflections. This is a manifestation of the symmetry of the bilayer's intrinsic structure. If, however, the cutout boundaries do not respect the bilayer's symmetry, e.g., III and IV of panel (b), strips with a well-defined handedness result, as seen in III and IV of panel (c). The handedness observed depends solely on the orientation of the strip's long axis; strips aligned with one diagonal generate right-handed helicoids, whereas strips oriented in the perpendicular direction generate left-handed helicoids. Slicing a narrow piece from a left-handed strip such that its aspect ratio is inverted yields a narrower strip of opposite handedness, as seen in V of panel (c), which was cut from VI. The square cutout II in panel (c) holds the capacity to generate both right- and left-handed strips. We thus consider it as possessing both right and left handedness in equal amounts rather than having no handedness. It is right handed along the $x$ direction and left handed along the $y$ direction. This directional dependence of the handedness is also observed in the relative positioning of cutouts II, III, and IV of panel (c), where the symmetric cutout II can be seamlessly continued into manifestly right- or left-handed helical structures. Such an orientation-dependent handedness cannot be captured by any pseudoscalar measure, and it calls for quantification by a pseudotensor.

Fig. 1(c)], as expected, gives rise to no distinct handedness. However, it is capable of manifesting extrinsic handedness when considered in specific directions, right along the $\pm x$ directions and left along the $\pm y$ directions, in agreement with the handedness of the elongated strips. In this way, handedness emerges from a combination of the intrinsic structure of the bilayer and a choice of direction.

Neither (pseudo)scalars nor (pseudo)vectors are capable of capturing the orientational behavior described above [20]. The simplest object that captures such an orientational variation is a (rank-2) pseudotensor, such as the one shown in Eq. (1):

$$\mathcal{X} = c \begin{pmatrix} 1 & 0 & 0 \\ 0 & -1 & 0 \\ 0 & 0 & 0 \end{pmatrix}. \tag{1}$$

This pseudotensor is symmetric under reflection, associates the $x$ direction with a right-handed $(+)$ rotation about the $x$ axis as observed in the $\chi_{xx}$ component, and associates the $y$ direction with a left-handed $(-)$ rotation about the $y$ axis as observed by the $\chi_{yy}$ component.

Furthermore, it may be easily shown that every rank-2 mirror-symmetric pseudotensor in three dimensions must have a vanishing determinant and trace [21]. It follows that if the pseudotensor is also symmetric, then it can be brought to the form of (1) by a rotation.

## III. INSTANCES OF HANDEDNESS PSEUDOTENSORS

A pseudotensor may be naturally interpreted as relating directions (vectors $\hat{\mathbf{n}}$) to rotations (pseudovectors $\mathbf{m}$) through $\mathbf{m} = \chi \hat{\mathbf{n}}$. In what follows, this is taken as the essential ingredient in constructing handedness measures. We note that the interpretation of handedness as a relation between directions and rotations is consistent with existing notions of handedness, e.g., the thread handedness for a screw, the right-hand rule, R/S classification of organic molecules, and optical activity.

In the following subsections, we construct explicit handedness measures following this principle. In each case, we begin by identifying the notions of direction and rotation that are related by the measure.

### A. Handedness pseudotensor for surfaces

In the case of embedded surfaces, we consider a handedness measure that arises from a local handedness density $\chi^e$. This handedness density pseudotensor is defined such that for every two unit vectors $\hat{\mathbf{n}}$ and $\hat{\mathbf{m}}$, we take the contraction $\hat{\mathbf{m}}\chi^e\hat{\mathbf{n}}$ to quantify the rotation of the surface's normal about the vector $\hat{\mathbf{m}}$ when it is displaced along the surface in the direction projected from $\hat{\mathbf{n}}$. It will be positive if the rotation about $\hat{\mathbf{m}}$ is right handed and negative when the associated rotation is left handed.

To formulate the above idea, we start by considering a surface $\mathbf{r}$ parametrized by the coordinates $x^\alpha$, where $\alpha = 1, 2$. These coordinates induce the metric $a_{\alpha\beta} = \partial_\alpha \mathbf{r} \cdot \partial_\beta \mathbf{r}$ and the second fundamental form $b_{\alpha\beta} = \partial_\alpha \partial_\beta \mathbf{r} \cdot \hat{\mathbf{N}}$, where $\hat{\mathbf{N}}$ is the surface's normal. Given a direction in space $\hat{\mathbf{n}}$ with Cartesian component $n_i$, we project it to the surface's tangent space by

$$\hat{\mathbf{n}}_\| = (\partial_\alpha \mathbf{r} \cdot \hat{\mathbf{n}}) a^{\alpha\beta} \partial_\alpha \mathbf{r}.$$

Differentiating a function $f$ defined on the surface along the projection of $\hat{\mathbf{n}}$ reduces to

$$(\hat{\mathbf{n}}_\| \cdot \nabla) f = (\partial_\beta \mathbf{r} \cdot \hat{\mathbf{n}}) a^{\alpha\beta} \partial_\alpha f.$$

For an oriented derivative of a vector field $(\hat{\mathbf{n}} \nabla) \mathbf{V}$, we may isolate the component that is due to a pure rotation about a vector $\hat{\mathbf{m}}$ by the scalar product $\hat{\mathbf{m}} \cdot (\hat{\mathbf{V}} \times (\hat{\mathbf{n}}\nabla)\mathbf{V})$. Whenever this product is positive, the change in the field $\mathbf{V}$ along $\hat{\mathbf{n}}$ is associated with a right-handed rotation about $\hat{\mathbf{m}}$. The rotation of the normal of a surface about a vector $\hat{\mathbf{m}}$ when displaced along the direction induced by the vector $\hat{\mathbf{n}}$ is then given by

$$\hat{\mathbf{m}} \cdot (\hat{\mathbf{N}} \times ((\partial_\beta \mathbf{r} \cdot \hat{\mathbf{n}}) a^{\alpha\beta} \partial_\alpha \hat{\mathbf{N}})) = \hat{\mathbf{m}} \chi^e \hat{\mathbf{n}}, \quad (2)$$

where the chirality density $\chi^e$ defined by Eq. (2) can be rewritten in component form as

$$\chi^e_{ij} = \partial_\alpha r_j a^{\alpha\beta} \epsilon_{ilk} N^l \partial_\beta N^k, \quad (3)$$

where $r_j$ and $N^k$ denote the Cartesian components of $\mathbf{r}$ and $\hat{\mathbf{N}}$, and $\epsilon$ is the antisymmetric Levi-Civita tensor. We may eliminate the normal vector from the formulation with the aid of the components of the second fundamental form $b_{\alpha\beta}$ and the two-dimensional Levi-Civita tensor $\epsilon^{\delta\gamma}$:

$$\chi^e_{ij} = \partial_\alpha r_j \partial_\delta r_i a^{\alpha\beta} b_{\gamma\beta} \epsilon^{\delta\gamma} / \sqrt{|a|}. \quad (4)$$

It is easy to show that the handedness density given above transforms as a pseudotensor and is independent of the surface's parametrization and of the sign of the normal vector.

This handedness density, similar in spirit to the tensorial measure proposed in Refs. [22,23] in the context of nematic ordering by chiral probes, may be integrated to give a tensorial handedness measure of the surface as a whole: $\mathcal{X}^e_{ij} = \iint \chi^e_{ij} dA$.

For example, calculating $\mathcal{X}^e$ for the symmetric cutout in part II of Fig. 1(c) yields a handedness tensor of the form given in Eq. (1), as expected from its symmetry, with $c = 14$ mm.

When the same measure applied above to a mirror-symmetric object is applied to the elongated helicoidal strips, parts IV and III, it gives rise to diagonal handedness tensors with the diagonal components $(88, -44, -44)$ mm and $(44, -88, 44)$ mm, respectively. These tensors are no longer mirror symmetric but are mirror images of each other; nonetheless, they are traceless. This is, in fact, a hallmark of the local mirror symmetry of two-dimensional surfaces. Every handedness density of the form $\chi_{ij} = \partial_i r_j \partial_\beta r_i \tilde{\chi}^{\alpha\beta}$, where $\tilde{\chi}^{\alpha\beta}$ is a function of the local surface properties defined only through the first and second fundamental forms, $a$ and $b$, must have a vanishing trace. For the measure presented above, an explicit proof of this property is given by

$$\chi^e_{ii} = a_{\alpha\delta} a^{\alpha\beta} b_{\gamma\beta} \epsilon^{\delta\gamma} / \sqrt{|a|} = b_{\gamma\delta} \epsilon^{\delta\gamma} / \sqrt{|a|} = 0,$$

where the last equality follows from the contraction of a symmetric and an antisymmetric tensor. As discussed in the next subsection, this property is unique to surfaces and does not follow for filamentous structures or when considering a three-dimensional director field.

## B. Handedness pseudotensor for 3D director fields

We now consider a unit vector field $\mathbf{u}$, such as that used to describe the director field of a nematic or cholesteric mesophase of a liquid crystal. In this case, we take the contraction $\hat{\mathbf{m}} \chi^e \hat{\mathbf{n}}$ to measure the rotation of the unit vector field $\mathbf{u}$ about the vector $\hat{\mathbf{m}}$ when displaced along the direction $\hat{\mathbf{n}}$. In components, this takes the form

$$n^i \chi_{ij} m^j = n^i \partial_i u^k \epsilon_{jlk} u^l m^j.$$

The trace of the handedness tensor defined above gives

$$\chi^{ii} = \partial_i u^k \epsilon_{ilk} u^l = -(\nabla \times \mathbf{u}) \cdot \mathbf{u},$$

which coincides with the expression for helicity, cf. magnetic helicity $[\mathbf{A} \cdot \mathbf{B} = \mathbf{A} \cdot (\nabla \times \mathbf{A})]$ and hydrodynamic helicity $[\mathbf{u} \cdot \omega = \mathbf{u} \cdot (\nabla \times \mathbf{u})]$. Note that as the handedness density is quadratic in the unit vector field $\mathbf{u}$, it remains unchanged under the transformation $\mathbf{u} \rightarrow -\mathbf{u}$, allowing us to interpret $\mathbf{u}$ as a director field. For example, a simple cholesteric order in which the director field is given by

$$\mathbf{u} = (\cos(pz), \sin(pz), 0)$$

displays a uniaxial handedness density oriented along the $z$ direction,

$$\chi = \begin{pmatrix} 0 & 0 & 0 \\ 0 & 0 & 0 \\ 0 & 0 & p \end{pmatrix}.$$

In general, when this handedness measure is applied to director fields, it not only yields the degree of handedness (such as the cholesteric pitch) but also associates the handed phenomena with a direction.

## C. Handedness pseudotensor for curves

Finally, we consider the handedness of embedded curves following the same guiding principles described above. In this case, the contraction $\hat{\mathbf{m}} \chi \hat{\mathbf{n}}$ gives the rotation of the curve's normal vector, **N**, about the direction $\hat{\mathbf{m}}$ when displaced along the curve in the direction and magnitude projected from $\hat{\mathbf{n}}$.

We define **t**, **N**, and **b** to be a curve's tangent vector, normal vector, and binormal vector, respectively. These unit vectors satisfy the Serret-Frenet formulas:

$$\partial_s \begin{pmatrix} \mathbf{t} \\ \mathbf{N} \\ \mathbf{b} \end{pmatrix} = \begin{pmatrix} 0 & \kappa & 0 \\ -\kappa & 0 & \tau \\ 0 & -\tau & 0 \end{pmatrix} \begin{pmatrix} \mathbf{t} \\ \mathbf{N} \\ \mathbf{b} \end{pmatrix},$$

where $s$ is the arc-length parametrization of the curve and $\kappa$ and $\tau$ are the curve's curvature and torsion. Differentiating along the curve in the direction projected from $\hat{\mathbf{n}}$ gives a weighted arc-length derivative $\mathbf{t} \cdot \hat{\mathbf{n}} \partial_s$. The handedness density tensor may be simplified by the Serret-Frenet equations to read

$$\chi_{ij} = t_i \epsilon_{jlk} N^l \partial_s N^k = t_i t_j \tau + t_i b_j \kappa. \quad (5)$$

The trace of the handedness density gives the local torsion $\chi_{ii} = \tau$. For locally planar curves where $\tau = 0$, this gives a traceless tensor, as expected from the local mirror symmetry of such curves. In the general case, however, the measure is not traceless. For example, when considering a helix oriented along the $z$ axis of pitch $p$,

$$\mathbf{r} = \left[ R \cos\left( \frac{s}{\sqrt{R^2 + p^2}} \right), R \sin\left( \frac{s}{\sqrt{R^2 + p^2}} \right), \frac{ps}{\sqrt{R^2 + p^2}} \right].$$

Then, for an integer number of windings, $M$, the integrated handedness tensor is uniaxial and oriented along the axis of the helix:

$$\mathcal{X} = \begin{pmatrix} 0 & 0 & 0 \\ 0 & 0 & 0 \\ 0 & 0 & \frac{2\pi M p}{\sqrt{R^2 + p^2}} \end{pmatrix} = \begin{pmatrix} 0 & 0 & 0 \\ 0 & 0 & 0 \\ 0 & 0 & \frac{\Delta Z}{\sqrt{R^2 + p^2}} \end{pmatrix},$$

where $\Delta Z$ is the height of the helix.

## IV. PROBING THE DIFFERENT COMPONENT OF HANDEDNESS

The examples above describe handedness measures that are geometric in nature and stem from a local handedness density. In the general case, handedness pseudotensors need not be of a purely geometric nature and may not be the result of linear summation from a local density. Nonetheless, the different tensor components may be probed operationally. An example of this is provided by the experiment reported in Fig. 2, which shows a section of a right-handed helicoid that was constrained to rotate about each of two perpendicular axes and subjected to an airflow along the constraint axis. When constrained along its longest direction, the flow induces a left-handed rotation, as expected. When constrained to rotate about the perpendicular direction, the same flow induced a (faster) *right*-handed rotation. While the handedness tensor describing high-Reynolds-number flow past an object does not arise from the summation of a local-handedness density, we note that some of the properties described above have counterparts in this setting. In particular, we observe that this surfacelike object displays two opposite notions of handedness along different directions.

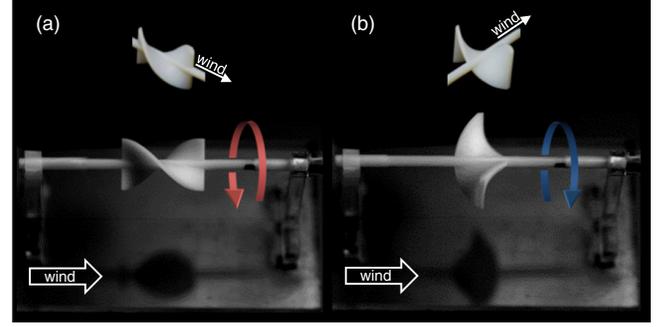

FIG. 2. Probing the different components of a handedness pseudotensor: Air flow past a helicoid. Two identical right-handed helicoidal surfaces, supported by thin cylindrical rods oriented along perpendicular directions, display opposite response to airflow. The helicoids were printed using a three-dimensional printer (Objet Connex350™), and they measure 2 cm wide and $\pi$ cm long. The axis of the first helicoid is oriented along its length (a), whereas the axis of the second helicoid is oriented along the transverse direction (b). The structures were placed in an airflow, and their axes were hinged to allow free rotation about the direction of the flow. (a) As air flows past the longitudinal axis of the helicoid, the latter rotates in a left-handed fashion. (b) The same helicoid rotates in a right-handed fashion when hinged along the perpendicular direction. Surprisingly, the helicoid hinged along the traverse direction rotates faster than the one hinged along its long axis.

Another setting in which orientation-dependent handed response is observed is optical activity [19,24]. Traditionally, optical activity has been used as a probe for the chiral shape of invisible molecules, the implicit assumption being the existence of a direct connection between a "chiral" electronic shape of each molecule, randomly oriented in solution, and the rotation of the polarization of light traveling through the solution. The manifestation of any handed phenomena in such isotropic collections of scatterers necessitates the absence of mirror symmetry and naturally gives rise to an isotropic handed response. For nonisotropic structures, however, optical activity can be displayed by mirror-symmetric objects and may depend on the relative orientation of the structure and incident light. This can be seen, for example, in the optical activity of a mirror-symmetric crystal [19], which displays equal amounts of left- and right-handed optical response in different directions, and in the nontrivial angular dependence of the optical activity displayed by an array of aligned right-handed helical copper wires [24]. In the next section, we examine how this orientation dependence can be exploited in the design of optically active metamaterials.

## V. OPTICAL ACTIVITY IN METAMATERIALS AND CHIRAL DESIGN

Modern fabrication techniques have recently enabled the production of large arrays of oriented conducting scatterers with prescribed geometric structures. While the scattering off any given geometric structure can be efficiently numerically simulated or experimentally measured, there is little in the way of a guiding principle available for rationalizing the design of scatterers to meet a desired optical response. While many studies have focused on chiral scatterers, such as helices [6], recent investigations of constructionally favorable planar structures etched out of a conducting surface have revealed a surprisingly strong response along with nontrivial angular dependence [4].

In what follows, we show how the tensorial structure of handed response, applied to objects possessing structural symmetries, can, with no additional information, be used to interpret the observed angular dependence and exploited to rationally design structures with enhanced optical performance.

We first address the orientational behavior of a single scatterer of the form considered in Ref. [4]. The planar structure automatically renders the scatterers mirror symmetric. As light can only rotate about its direction of propagation, the relevant handedness tensor is necessarily symmetric. Thus, as discussed above, mirror symmetry implies that the rank-2 handedness response must have the form of Eq. (1). This allows us to predict the orientational variation of the handed optical response without knowledge of the specific details of the scattering process, and to collapse the different response curves obtained in Ref. [4] by examining different orientations onto one master curve, with no fitting parameters. Thus, the entire orientational response in this case can be captured by a single number that characterizes the maximal response.

The symmetry of the object and the assumption that the response is characterized by a tensor are the only ingredients that are required in the reasoning given above. We therefore expect the same reasoning to hold for any object possessing the same symmetries. This can be observed in Fig. 3(a), where we show the numerically computed orientation-dependent optical activity of a semicircular scatterer possessing the same symmetry as the structure considered in Ref. [4]. The scatterer's structure is symmetric under reflection (about two perpendicular planes) and planar, rendering it also similar to bent core molecules that form polar nematic phases and were recently shown to exhibit an isotropic chiral liquid crystal phase [16]. As expected, to within small contributions of higher-order tensors (whose orientational variation is calculated explicitly in Appendix B), the optical activity obeys the form of Eq. (1). This form predicts an angular dependence that scales with $\sin(2\alpha)$, where $\alpha$ is the angle between the normal to the scatterer plane and the $k$ vector and rescaling by this factor yields the collapse observed in Fig. 3(a).

This approach has immediate applications in the design of metamaterials to achieve a desired optical response, such as, for example, a purely optically active metamaterial. Precise control of the polarization response of a metamaterial requires engineering not only the optical activity (circular birefringence) but also the linear birefringence, which is often comparable or greater in magnitude. In the case of the semicircular scatterer considered above, the optical activity and linear birefringence effects are roughly equal in magnitude [Fig. 3(b)]. The desired polarization response of a given structure may often be found at a single frequency, where contributions of multiple resonances add up favorably. For example, in Ref. [4], attenuation of linear birefringence (pure optical activity) was achieved by examining the resonant response of the metamaterial and identifying a single frequency in which linear birefringence is suppressed while the optical activity is maintained. An alternative, more geometric route follows from the distinct transformation rules implied by the pseudotensorial structure of the optical activity response. While linear birefringence is expected to change sign when the scatterer is rotated by $\pi/2$ about the $k$ vector direction, the optical activity is expected to remain unchanged by this rotation. To diminish the linear birefringence response in all directions, one may attempt to impose this symmetry in all directions. However, this also results in the vanishing of the optical activity pseudotensor. By contrast, isotropically orienting the scatterer's polar vector and director separately (as done in Fig. 3) is less restrictive and allows for a constructive handed response. We implement this principle to design a spatial arrangement of six semicircular scatterers that attenuates the linear birefringence while retaining the optical activity [Fig. 3(b)]. This arrangement constitutes a unit cell that, similarly to its semicircular building blocks, is mirror symmetric, making it possible to use the same principle to construct a hierarchal bulk of scatterers.

The essence of the approach presented above is to exploit the difference in the transformation rules of tensors and pseudotensors that quantify the effective circular and linear optical effects. Such an approach requires little knowledge of the excitations of the scatterers that give rise to the specific effects, or any fine-tuning of resonances, as the transformation rules are independent of such details. Thus, associating the optical activity of the scatterers with a handedness tensor (relating directions to rotations) naturally leads to a principle of frequency-independent geometric design that complements existing schemes and approaches (see Appendix C, and Refs. [13,25]).

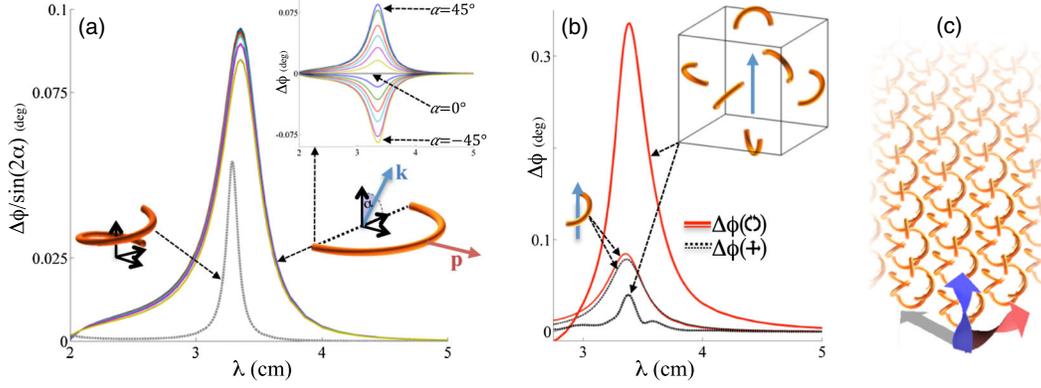

FIG. 3. Optical activity and the design of an optically active mirror-symmetric array of conducting scatterers. (a) The optical activity (relative phase delay between right and left circular polarizations) of a semicircular conducting wire calculated numerically at varying angles of incidence and plotted against the exciting wavelength (see Appendix A). The inset shows the rotation angle $\Delta\phi$ as a function of the exciting frequency, where different curve colors correspond to different orientations, as described below. The main panel shows the collapse of the rotation angle curves when rescaled by $\sin(2\alpha)$. The peak response is observed at a wavelength $\lambda/2 \sim l$, where $l \sim \pi/2$ cm is the length of the scattering wire. The orientation of the semicircles is prescribed by a director (axis) that joins the semicircle's ends (dotted line) and a polar vector **p** perpendicular to the director and pointing to the midpoint of the semicircle. The angle of incidence, $\alpha$, is measured between the incident **k** vector and the normal to the semicircle. $\alpha$ is varied by rotating the semicircle about its polar vector, and it takes the values $\pm 45$, $\pm 30$, $\pm 20$, $\pm 15$, $\pm 10$, $\pm 5$, and 0. The strong angular dependence of the response amplitude (inset) is predicted by (1) to scale as $\sin(2\alpha)$ at all wavelengths, as observed by the collapse of the rescaled curves. The optical activity of a single period of a helix of the same length is given for comparison (dashed curve). The small discrepancy between the rescaled curves is due to a small fourth-order tensor correction, which can be calculated explicitly. See Sec. VI. (b) To reduce linear birefringence of a single scatterer (thin black curve) while retaining its optical activity (thin red curve), we make use of the tensorial form of the optical activity (see Table I, the symmetry arguments). This allows a mirror-symmetric arrangement of six semicircles on the faces of a cube such that both their directors and polar vectors are isotropically distributed while retaining a constructive handed response. The relative magnitude between the resulting optical activity (thick red curve) and the linear birefringence (thick black curve) is reduced by a factor of about 8. As the handedness tensor of the array of scatterers possesses the same symmetry properties as that of the single scatterer, an array of such arrays may be constructed to further diminish linear effects. Successive iteration will result in a hierarchical arrangement of scatterers exhibiting an exponential (in the iteration number) attenuation of linear birefringence and dichroism effects. (c) A section from an optically active metamaterial sheet, where each scattering unit is made from six semicircular scatterers arranged as described in (b). The helical arrows denote the principal directions of optical activity.

## VI. HIGHER-RANK HANDEDNESS PSEUDOTENSORS

The quantifications of handedness in the examples considered so far were dominated by rank-2 pseudotensors. In some cases, such a quantification is exact. For example, in viscous flows, because of the linearity of the governing equations, there is a general linear relation between the velocity $U$ and rotation $\Omega$ of a body in an otherwise quiescent viscous fluid and the force $F$ and couple $G$ applied to it through a resistance matrix [26]:

$$\begin{pmatrix} F \\ G \end{pmatrix} = \begin{pmatrix} A & L \\ L^T & B \end{pmatrix} \begin{pmatrix} U \\ \Omega \end{pmatrix}.$$

The off-diagonal block $L$ relating a given rotation to a linear force, as well as the combination $B^{-1}L^T$ that relates the rotation due to translation of a torque-free body, is a measure of handedness that is captured exactly by a rank-2 tensor.

However, adopting the interpretation of handedness as a relation between directions and rotations does not necessarily imply that this relation is linear, and in general, the handedness response admits more complicated orientational structure. In such cases, the handedness function that relates the direction $\hat{\mathbf{r}}$ with the rotation about $\hat{\mathbf{n}}$, $\chi(\hat{\mathbf{r}}, \hat{\mathbf{n}})$, may be expanded in a formal tensor series

$$\chi(\hat{\mathbf{n}}, \hat{\mathbf{r}}) = r^i \chi^{1,1}_{ij} n_j + r_i r_j \chi^{2,1}_{ijk} n_k + r_i r_j r_k \chi^{3,1}_{ijkm} n_m \\ + r_i \chi^{1,3}_{ijkm} n_j n_k n_m + \cdots.$$

It is important to note that in some cases symmetry sets the second-rank tensor to zero, and the dominant behavior is governed by a higher-rank tensor. This is the case of the four-cup anemometer that is designed to spin about the $z$ axis in the same fashion for wind both from the $x$ direction and from the $-x$ direction. The lowest order pseudotensor that is capable of such a response is the rank-3 $\chi^{2,1}$.

Even in cases where the second-rank contribution dominates the orientational variation of the handedness, we may observe the effect of the higher-rank corrections. The collapsed curves of Fig. 3(a) show some variation ($\lesssim 10\%$), which is especially discernible near the resonant frequency. This variation can be attributed to the higher-order corrections and treated much in the same manner that we treated the rank-2 contribution exploiting the symmetry of the scatterer. In Appendix B, we explicitly calculate the third- and fourth-rank contributions to the orientational variation, yielding that the former vanishes identically while the latter yields a variation that scales as $\cos(2\alpha)$. When these corrections are applied to the collapsed curves, the variation is reduced to less than 1%.

## VII. RELATING HANDEDNESS AND CHIRALITY

A distinction between handedness, which resolves right from left, and chirality, which is associated with the lack of mirror symmetry, was first made formal by Ruch [11]. According to Ruch, handedness is identified as a property that may be unambiguously attributed only to a subclass of objects. In particular, the handedness of an object is a pseudoscalar, and all handed objects are chiral but not vice versa. In the present work, we take a different path, defining handedness as a relation between directions and rotations. This leads to a different relation between handedness and chirality, which we review next.

The quantification of handedness defined as a relation between directions and rotations is naturally orientation dependent and thus tensorial. Unlike the case for pseudoscalars, a nonvanishing pseudotensor may display mirror symmetry, as shown in Eq. (1). Moreover, the definition of handedness as a relation between directions and rotations sidesteps the difficulties arising from chiral connectedness, as discussed in the Introduction, because the lack of mirror symmetry is not taken to be the source of handedness (see Appendix D for further discussion).

This approach also accounts for the fact that determining the handedness of some objects with a well-defined axis, such as a screw or a helix, is a relatively easy task: The objects' natural axis allows us to apply some variant of the right-hand rule, which is equivalent to singling out one component of the appropriate handedness tensor. If the same criteria for handedness are applied to perpendicular directions, the remaining components of the handedness tensor may be obtained.

Upon taking the isotropic average of handedness tensors, which is physically equivalent to computing the average handed response of a collection of randomly oriented copies of the object, as might occur for molecules in a solution, we recover a pseudoscalar (the trace) that is consistent with Kelvin's definition; for the isotropic average handedness of an object to be nonvanishing, the object must break mirror symmetry. However, for single objects, or aligned collections of objects, tensorial measures predict handed behavior that depends on their orientation. We suggest that such tensorial measures provide a natural extension for the definition of handedness, where Kelvin's definition of chirality constitutes the isotropic contribution to handedness.

## VIII. CONCLUSIONS

The handed response of nonisotropic objects may vary in both magnitude and sign when examined along different directions. Attributing this varying response to the object itself leads to an orientation-dependent quantification of handedness that sidesteps the difficulties commonly associated with quantification of chirality. The resulting tensorial handedness measure also allows a simple interpretation relating directions to rotations. We have demonstrated the predictive power of this approach in settings ranging from purely geometric measures of the structure of fields and objects to the optical response of electromagnetic metamaterials. In some settings, this quantification leads to nonintuitive results such as the inevitable left handedness (in certain directions) of a right-handed helicoidal surface with respect to appropriate handedness measures.

The tensorial quantification of handedness also opens new avenues for the design of chiral objects and the manipulation of their chiral response. This gain, however, does not come without a cost; the handedness of a general object can no longer be simply stated by a single identifier, e.g., "right handed," and an oriented statement such as "isotropically left handed" or "uniaxially right handed" take its place. In the most general case, one must resort to the use of tensors to fully capture the handedness of an object.

## ACKNOWLEDGMENTS

We thank P. M. Chaikin, N. Clark, H. Jaeger, R. Kamien, D. Kleckner, C. Santangelo, J. Selinger, H. K. Moffatt, and T. Witten for stimulating and helpful discussions. We acknowledge the support of the University of Chicago MRSEC under NSF Grant No. DMR-0820054. E. E. additionally acknowledges the support of the Simons Foundation. W. T. M. I. further acknowledges support from the A. P. Sloan Foundation and the Packard Foundation.

## APPENDIX A: METHODS

### 1. Construction and geometry of the elastic bilayer

Two sheets of a supersoft silicone rubber (30a Durometer) of thickness $1/32$ in were strained uniaxially by 25% and their edges temporarily fixed to rigid plates. A thin layer of silicone adhesive (Dow Corning 732) was applied to both sheets. The sheets were then pressed against each other while maintaining the perpendicular uniaxial strain axes. The silicone was allowed to cure for a week. The glued bilayer shows a thickness of 2 mm with a spatial variation of ±20% when released from the tensional frame.

Following Ref. [27], we identify the intrinsic surface geometry resulting from the construction of the bilayer with uniform first and second reference fundamental forms:

$$a = \begin{pmatrix} 1 & 0 \\ 0 & 1 \end{pmatrix}, \qquad b = \begin{pmatrix} 0 & k \\ k & 0 \end{pmatrix},$$

where principal curvature directions coincide with the tension directions in each of the layers. The magnitude of the reference curvature may be calculated via $k = \frac{1}{\delta z} \frac{\alpha^2 - 1}{\alpha^2 + 1}$, where $\alpha$ is the uniaxial elongation factor with respect to the unstrained state and reads $\alpha = 1.25$, and $\delta z \approx 1.4$ mm is measured between the midsurfaces of the top and bottom layers. The resulting curvature reads $k \approx 0.15 \frac{1}{mm}$.

### 2. Three-dimensional printed helicoid segments in an airflow

Half-period helicoids of width 2 cm, length $\pi$ cm, and thickness 0.2 cm constrained by 0.4-cm-diameter and 11-cm-long cylindrical axes were printed using a 3D printer (Objet Connex350™). The constraint axes passed through the helicoid's center and were directed in perpendicular directions, along the helicoid's length and tangent to the helicoid at its center.

### 3. Optical activity simulations

We used a commercial method of moments (MoM) solver, *HFSS-IE*, to solve the near-field scattering off a conducting solid (copper from the *HFSS* built-in material library). The scatterer was centered at the origin, which was chosen as the zero phase of the incident plane-wave excitation. The exciting wave's wavelength was varied between 10 mm and 150 mm, revealing a single resonance approximately at the scatterer's length. The semicircular ring of Fig. 3 is one-half of a torus of minor radius 0.05 mm and a major radius of 5.05 mm, the helical segment is of the same wire thickness (0.05 mm), and a radius and pitch $r_0 = p \approx 2.4$ mm, resulting in the same overall wire length. A near-field measurement at a distance of 2000 mm from the origin of the scattered field was added to the unattenuated incident plane wave. Two linearly polarized incident-wave calculations were used to obtain the different components of the Jones matrix. The linear and circular components of the birefringence were extracted under the assumption of a homogenous media, as elaborated in following subsection.

### 4. Optical activity extraction from the Jones matrix

Monochromatic scattering is described by a Jones matrix relating the incident and scattered plane-wave components:

$$\begin{pmatrix} E_x^{out} \\ E_y^{out} \end{pmatrix} = \begin{pmatrix} J_{xx} & J_{xy} \\ J_{yx} & J_{yy} \end{pmatrix} \begin{pmatrix} E_x^{in} \\ E_y^{in} \end{pmatrix}.$$

Following Jones [28], we identify the four independent complex entries of the Jones matrix, $J$, with the eight independent polarization transformations: (i) a global phase retardation and amplitude attenuation, (ii) circular dichroism and circular birefringence, (iii) linear birefringence and linear dichroism in the $(x, y)$ linear polarizations, and (iv) linear birefringence and linear dichroism in the $(x + y, x - y)$ linear polarizations. As the different Jones matrices associated with the different polarization phenomena do not commute, a decomposition of a Jones matrix to a product of such "pure" components will not be unique, and the values associated with the strength of each phenomenon will depend on the order of the elements in the product. If, however, we consider the generators of the polarization effects, then to linear order, the result does not depend on the order in which they are summed. Conceptually, this amounts to decomposing the medium to infinitesimal laminae, each displaying one of the polarization effects. As their thickness is infinitesimal, the order in which these laminae are placed one after the other does not change the result. This assumption of a homogenous medium results in the following interpretation of the Jones matrix components.

$$J = e^{-i\eta/2} \begin{pmatrix} \cos\left(\frac{T}{2}\right) - i\frac{L}{T}\sin\left(\frac{T}{2}\right) & \frac{C - i\tilde{L}}{T}\sin\left(\frac{T}{2}\right) \\ -\frac{C + i\tilde{L}}{T}\sin\left(\frac{T}{2}\right) & \cos\left(\frac{T}{2}\right) + i\frac{L}{T}\sin\left(\frac{T}{2}\right) \end{pmatrix},$$

where $T = \sqrt{L^2 + \tilde{L}^2 + C^2}$, $L = LB_{0,90} - iLD_{0,90}$ measures the linear birefringence and dichroism of the linear polarization along the $x$ and $y$ axes, $\tilde{L} = LB_{45,-45} - iLD_{45,-45}$ measures the linear birefringence and dichroism of the linear polarization along the $x + y$ and $x - y$ directions, $C = CB - iCD$ measures the circular birefringence and dichroism, and $\eta$ accounts for the isotropic amplitude attenuation and phase retardation [28,29]. For a weak scatterer, the extraction of the optical activity from the Jones matrix, $J$, is unique.

## APPENDIX B: ORIENTATIONAL VARIATION OF THE OPTICAL ACTIVITY OF A SINGLE SCATTERER

We compute the corrections up to rank 4 to the optical activity of the single scatterer considered in Fig. 3 of the main text. As in the case of the rank-2 tensor for a mirror-symmetric object, symmetries serve to reduce the number of independent entries.

For the rank-3 tensor, we obtain that the nonvanishing components are $\chi^{2,1}_{\{123\}}$ and $\chi^{2,1}_{\{132\}}$, where {123} denotes all possible cyclic index permutations, i.e., [(123), (231), (312)]. For the rank-4 tensor, the only nonvanishing components are

$$\chi^{3,1}_{\{1113\}}, \qquad \chi^{3,1}_{\{3331\}}, \qquad \chi^{3,1}_{\{2123\}}, \qquad \chi^{3,1}_{\{2231\}}, \qquad \chi^{3,1}_{\{2213\}}.$$

We note that for optical activity calculations, where the **k** vector serves as both the direction of propagation and the axis of rotation, $\chi^{1,3}$ and $\chi^{3,1}$ relate to the same tensor; thus, only one of them needs to be prescribed. In particular, when contracted with a single vector $\mathbf{k} = (\cos(\alpha), 0, \sin(\alpha))$, we obtain

$$k^i k^j \chi^{1,1}_{ij} = \cos(\alpha)\sin(\alpha)(\chi^0_{13} + \chi^0_{31}),$$

$$k^i k^j k^l \chi^{2,1}_{ijl} = 0,$$

$$\begin{aligned}
k^i k^j k^l k^n \chi^{3,1}_{ijln} &= \cos(\alpha)\sin(\alpha) \\
&\times (\cos^2(\alpha)(\chi^{3,1}_{1333} + \chi^{3,1}_{3133} + \chi^{3,1}_{3313} + \chi^{3,1}_{3331}) \\
&+ \sin^2(\alpha)(\chi^{3,1}_{3111} + \chi^{3,1}_{1311} + \chi^{3,1}_{1131} + \chi^{3,1}_{1113})) \\
&= \cos(\alpha)\sin(\alpha)(A_1 \cos(2\alpha) + A_2), \quad (B1)
\end{aligned}$$

where $A_1$ and $A_2$ are independent of $\alpha$. Subtracting the curves rescaled by $\cos(\alpha)\sin(\alpha)$ (observed in Fig. 3 in the main text) from the curve of maximal response results in the curves observed in Fig. 4, which scale as $\cos(2\alpha)$. This collapse also displays some variation that may be attributed to even higher-order corrections.

We note that, in general, $\chi^{2,1}$ need not vanish. In particular, for planar scatterers of lesser symmetry, such as those that appear in Ref. [30], we expect the dominant correction to result from this rank-3 contribution. For example, the first nonvanishing contribution to optical activity in normal incidence to planar metamaterial is predicted to be given by $\chi_{zzz}$. This result, in turn, predicts that additional reflectional symmetries in the plane will eliminate this effect and that forward and backward scattering will have opposite signs, consistent with the observed results and other symmetry-based arguments.

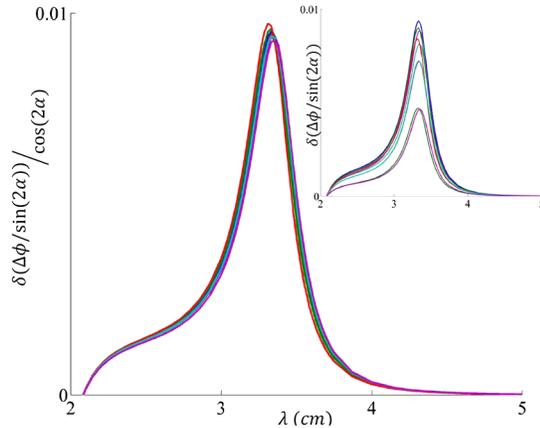

FIG. 4. Rank-4 tensor corrections to the optical activity. The difference between the collapsed curves in Fig. 3 of the main text is observed mostly around the resonant peak, and it is about an order of magnitude smaller than the peak value (inset). Further calculation shows that the symmetry of the problem allows no nontrivial rank3 pseudotensors and restricts the form of the rank-4 tensors to display a $\cos(2\alpha)$ angular dependence, as observed by the collapse of the rescaled curves.

## APPENDIX C: MAGNETOELECTRIC POLARIZATION TENSOR OF THE SPLIT RING ASSEMBLY

Recently, planar split-ring scatterers, such as the one in Fig. 3, were studied theoretically in terms of an electric-dipole and a magnetic-dipole coupled response. These studies focused on the optimization of the coupling of the scatterer to circularly polarized light. They presented an upper bound for this coupling strength [31] and showed that this bound is realized by every planar scatterer that can be considered as a single resonance circuit [25]. Generally, these scatterers also display non-negligible linear polarization effects. Eliminating the linear polarization effects for such scatterers requires that the magnetic polarizability be equal to the electric polarizability. Such cases are scarce and typically occur over a narrow frequency range. In this appendix, we show that the approach presented in this work and described in Fig. 3 allows a geometric path to eliminate all linear polarization effects in the scatterers described in Refs. [25,31], at all frequencies.

We next analyze the single planar scatterer and the array of scatterers shown in Fig. 3, employing the same units and conventions used in Ref. [31]; the single scatterer is taken to lie in the $xy$ plane, with the line connecting its ends aligned with the $x$ axis, and

$$\mathbf{E} = \mathbf{E}_{SI}, \quad \mathbf{H} = \frac{\mathbf{H}_{SI}}{\epsilon_0 c}, \quad \mathbf{p} = \frac{\mathbf{p}_{SI}}{4\pi\epsilon_0}, \quad \mathbf{m} = \frac{\mathbf{m}_{SI}}{4\pi\epsilon_0 c}.$$

The central tool in the analysis presented in Refs. [25,31] is the polarization tensor $\alpha$, defined through

$$\begin{pmatrix} \mathbf{p} \\ \mathbf{m} \end{pmatrix} = \alpha \begin{pmatrix} E_{in} \\ H_{in} \end{pmatrix}, \qquad \alpha = \begin{pmatrix} \alpha_{EE} & \alpha_{EH} \\ \alpha_{HE} & \alpha_{HH} \end{pmatrix},$$

where **p** and **m** are the excited electric and magnetic dipoles. For a planar split ring, the polarization tensor reads

$$\alpha = \begin{pmatrix} \eta_E & 0 & 0 & 0 & 0 & i\eta_C \\ 0 & 0 & 0 & 0 & 0 & 0 \\ 0 & 0 & 0 & 0 & 0 & 0 \\ 0 & 0 & 0 & 0 & 0 & 0 \\ 0 & 0 & 0 & 0 & 0 & 0 \\ -i\eta_C & 0 & 0 & 0 & 0 & \eta_H \end{pmatrix}.$$

Analysis of the eigenvalues of the polarization tensor performed in Ref. [31] yields that energy conservation places a bound for the cross coupling term $\eta_C \leq \sqrt{\eta_E \eta_H}$. This bound is surprisingly saturated by *every* planar scatterer that can be considered a single resonance circuit (where the magnetic and electric dipoles arise from the same equations of motion for the charge in the scatterer) [25]. However, only if $\eta_E = \eta_H$ does the system exhibit purely circular optical response and display no linear polarization effects.

To examine the different types of polarization effects, we construct a stripped-down version of the Green's function for the far fields generated by the electric and magnetic dipoles:

$$ G = \begin{pmatrix} G^{EE} & G^{EH} \\ G^{HE} & G^{HH} \end{pmatrix}, $$

where $G_{ij}^{EE} = G_{ij}^{HH} = \delta_{ij} - n_i n_j$ and $G_{ij}^{EH} = -G_{ij}^{HE} = \epsilon_{ijl} n_l$. We can then extract the relation between the scattered fields and incident fields from the tensor $G\alpha$. We examine the collimated response to a plane wave propagating along the $(1, 0, 1)/\sqrt{2}$ direction in which the maximal response is obtained [this is predicted both by the symmetry of the handedness tensor (1) and through explicit calculation [25]]. We express the result in terms of a Jones matrix written with respect to right-handed (+) and left handed (−) circularly polarized plane waves:

$$ J = \begin{pmatrix} T_{++} & T_{+-} \\ T_{-+} & T_{--} \end{pmatrix}. $$

Circular dichroism and birefringence arise from differences between the diagonal terms, whereas the off-diagonal terms give rise to linear polarization effects (dichroism and birefringence). For the single scatterer, we have

$$ J_{single} = \frac{1}{2} \begin{pmatrix} \eta_E + \eta_H + 2\eta_C & \eta_H - \eta_E \\ \eta_H - \eta_E & \eta_E + \eta_H - 2\eta_C \end{pmatrix}. $$

Note that $\eta_C$ alone gives rise to the circular polarization effects. However, even if we assume that the cross coupling term is optimal, in general, the off-diagonal terms are nonzero.

The arrangement described in Table I, obtained through the application of the tensor transformation rules, implies that one can arrange multiple scatterers such that their electric dipoles and magnetic dipoles are isotropically distributed, yet their handed optical effects add constructively. Arranging six scatterers according to Table I gives rise to isotropic electric and magnetic polarization tensors:

$$ \alpha = \begin{pmatrix} 2\eta_E & 0 & 0 & 0 & 0 & 2i\eta_C \\ 0 & 2\eta_E & 0 & 0 & 0 & 0 \\ 0 & 0 & 2\eta_E & 2i\eta_C & 0 & 0 \\ 0 & 0 & -2i\eta_C & 2\eta_H & 0 & 0 \\ 0 & 0 & 0 & 0 & 2\eta_H & 0 \\ -2i\eta_C & 0 & 0 & 0 & 0 & 2\eta_H \end{pmatrix}. $$

The Jones matrix for this arrangement reads

$$ J_{Array} = 4 \begin{pmatrix} \eta_E + \eta_H + \eta_C & 0 \\ 0 & \eta_E + \eta_H - \eta_C \end{pmatrix}, $$

which completely annihilates all linear birefringent and dichroic effects. Lastly, we note that the maximal circular polarization effects per scatterer read only 2/3 of the value for a single scatterer. Therefore, the result of this arrangement should be interpreted as selective attenuation of undesired effects rather than an increase in the response to circular polarization. Moreover, the tensorial sum given by Table I implies that the handed optical effects of the scatterer array display the same orientational dependence as those of a single scatterer. Thus, we expect that the result above is also valid at all directions and not only in the direction of maximal response. As this result is of a geometric origin, it is crucially independent of frequency.

TABLE I. We describe the orientation of a semicircular scatterer in space using a vector **p**, pointing from the center of the semicircle to the midpoint on the wire segment, and a director **d** (nonoriented vector, i.e., **d** = −**d**), which connects the segment's endpoints. Isotropically orienting **p** and **d** eliminates the optical activity $\chi$, together with the linear birefringence. However, exploiting the different transformation rules that **p**, **d**, and $\chi$ satisfy, we construct an arrangement of six scatterers on the faces of a cube such that their polar vectors **p** are isotropically oriented and their directors **d** are also isotropically oriented, but such that their handedness tensors add constructively. Moreover, this arrangement can be made to be symmetric under reflection; i.e., there exists an improper rotation $T$ (a combination of an inversion and a rotation) that maps the arrangement of scatterers to itself. The transformation $T$, mapping the scatterers' arrangement to itself, is given explicitly by $x \rightarrow -x$, $y \rightarrow -z$, and $z \rightarrow y$. The table below lists the spatial location **r**, polar vector **p**, director **d**, and the nonvanishing components of the handedness tensors $\chi$ for the optically active and reflection symmetric arrangement of scatterers, along with the values these quantities assume under the transformation described by $T$.

| **r** | **p** | **d** | $(\chi_{xx}, \chi_{yy}, \chi_{zz})$ | $T(\chi_{xx}, \chi_{yy}, \chi_{zz})$ | $T\mathbf{p}$ | $T\mathbf{d}$ | $T\mathbf{r}$ |
|---|---|---|---|---|---|---|---|
| $(L, 0, 0)$ | **x** | **y** + **z** | $(0, 1, -1)$ | $(0, 1, -1)$ | −**x** | **y** − **z** | $(-L, 0, 0)$ |
| $(-L, 0, 0)$ | −**x** | **y** − **z** | $(0, 1, -1)$ | $(0, 1, -1)$ | **x** | **y** + **z** | $(L, 0, 0)$ |
| $(0, L, 0)$ | **y** | **x** − **z** | $(1, 0, -1)$ | $(-1, 1, 0)$ | −**z** | **x** + **y** | $(0, 0, -L)$ |
| $(0, 0, -L)$ | −**z** | **x** + **y** | $(-1, 1, 0)$ | $(1, 0, -1)$ | −**y** | **x** + **z** | $(0, -L, 0)$ |
| $(0, -L, 0)$ | −**y** | **x** + **z** | $(1, 0, -1)$ | $(-1, 1, 0)$ | **z** | **x** − **y** | $(0, 0, L)$ |
| $(0, 0, L)$ | **z** | **x** − **y** | $(-1, 1, 0)$ | $(1, 0, -1)$ | **y** | **x** − **z** | $(0, L, 0)$ |

## APPENDIX D: ADDRESSING THE PARADOX OF CHIRAL CONNECTEDNESS

Chiral connectedness, the ability to continuously deform an object into its mirror image without passing through a configuration that is mirror symmetric, stands at the heart of the inability to quantify handedness in a manner consistent with Lord Kelvin's definition of chirality. The tensorial generalization to the quantification of handedness sidesteps this seeming contradiction. The oriented quantification of handedness proposed here does not take the lack of mirror symmetry to be the source of handedness but instead defines handedness as the relation between directions and rotations. As such, even non-mirror-symmetric objects are allowed to display no handedness with respect to a specific given measure.

It is important to state that even if this definition is not adopted, the tensorial quantification by itself relaxes the paradox of chiral connectedness. A higher-dimensional object (such as a tensor) can change sign without passing through zero, as can be seen in Fig. 5. This notion, in the context of vectors, was discussed by Weinberg and Mislow soon after the discovery of chiral connectedness, yet was discarded claiming that vectors do not allow simple ordering and thus cannot form adequate handedness measures [10].

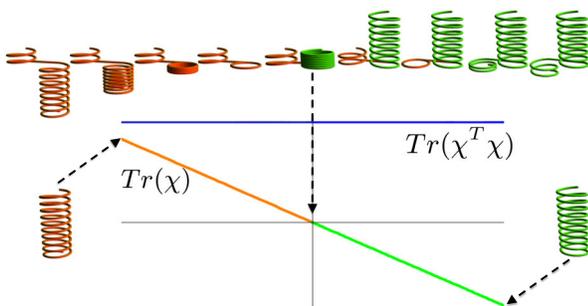

FIG. 5. Chiral-connected path between a right-handed helix and its left-handed mirror image. The first two isotropic invariants of the handedness tensor [calculated explicitly for the curve using the formula $\mathcal{X} = \int (\kappa t_i t_j + \tau t_i b_j) ds$ derived in Sec. III C] plotted against a parameter describing the continuous deformation of the right-handed helix (orange) to the left-handed helix (green). The first isotropic invariant $\chi_{ii}$ given by the green and orange line coincides with the scalar definition of chirality indices. This measure reads zero when the right-handed portion of the helix and the left-handed portion are the same height. However, this configuration is not symmetric under reflections, as the right-handed portion is composed of two windings, whereas the left-handed portion is composed of eight windings. The second invariant, $\chi_{ij}\chi_{ij}$ (divided by 10 for graphical purposes), is plotted in blue and barely changes throughout the transformation. In particular, it remains bound away from zero.

———